\documentclass[aip,author-year]{revtex4-1}

\usepackage{amsbsy}
\usepackage{amssymb}
\usepackage{amsmath}
\usepackage{graphicx}
\usepackage{color}

\def\aap{A \& A}
\def\apj{Ap. J.}
\def\apjl{Ap. J.}
\def\apjs{Ap. J. Supp. Ser.}
\def\mnras{MNRAS}

\def\prd{Phys. Rev. D}

\usepackage{psfrag}

\newcommand{\bxi}{\boldsymbol\xi}

\newcommand{\bv}{{\boldsymbol v}}

\newcommand{\beq}{\begin{equation}}
\newcommand{\eeq}{\end{equation}}
\newcommand{\bay}{\begin{array}}
\newcommand{\eay}{\end{array}}
\newcommand{\beqa}{\begin{align}}
\newcommand{\eeqa}{\end{align}}
\newcommand{\beqy}{\begin{eqnarray}}
\newcommand{\eeqy}{\end{eqnarray}}
\newcommand{\nn}{\nonumber}
\newcommand{\rmd}{\mathrm{d}}
\newcommand{\brac}[1]{\left({#1}\right)}

\newcommand{\td}[2]{\frac{\rmd{#1}}{\rmd{#2}}}

\renewcommand{\div}{\nabla\cdot}

\newcommand{\bB}{{\bf B}}

\newcommand{\be}{{\bf e}}

\newcommand{\br}{{\bf r}}

\newcommand{\Mr}{{\textstyle{\frac{\pi r}{R_*}}}}

\begin{document}

\title{The early life of millisecond magnetars}

\author{D. I. Jones}

\affiliation{Mathematical Sciences and STAG Research Centre, University of Southampton, Southampton SO17 1BJ, U.K.}

\begin{abstract}

Some neutron stars may be born spinning fast and with strong magnetic fields---the so-called \emph{millisecond magnetars}.  It is important to understand how a star's magnetic axis moves with respect to the spin axis in the star's early life, as this effects both electromagnetic and gravitational wave emission.  Previous studies have highlighted the importance of viscous dissipation within the star in this process.  We advance this program by additionally considering the effect of the electromagnetic torque. We find an interesting interplay between the viscous dissipation, which makes the magnetic axis orthogonalise with respect to the spin, verses magnetic torques that tend to make the magnetic axis align with the spin axis.  We present some results, and highlight areas where our model needs to be made more realistic.

\end{abstract} 
  
\maketitle


\section{Introduction}\label{sect:intro} 

In this paper we describe recent efforts to describe how the angle between the spin axis and the magnetic axis evolves, early in the life of a millisecond magnetar.  It is based upon two recent  papers \citet{LJ17, LJ18}, and also more recent Lander and Jones work.  

This problem is of interest for several reasons.  Firstly, millisecond magnetars have been considered as possible sources of gravitational waves, but only if the magnetic  axis is misaligned with the spin axis \citep{dallosso09, dallosso_spin, cutler_02}.  Secondly, there have been claims that the injection of energy from a rapidly spinning-down magnetar may partially power some gamma-ray bursts \citep{metzger}.  Finally, the results of our analysis could be compared against the observed distribution of inclination angles in the observed Galactic magnetar population.  We do not discuss these interesting issues in what follows, but rather describe our attempts thus far to build a model of a spinning highly magnetised star that contains many (but not yet all) of the important ingredients in this problem.

\section{Precession of a magnetised star}\label{sect:poams}

We model our star as consisting of a spherically symmetric background configuration, with density $\rho_0(r)$, on which is superimposed two perturbations.  One is due to the magnetic field, and produces a density perturbation $\delta\rho_B(r,\theta)$, while the other is due to rotation, and produces a density perturbation $\delta\rho_\alpha(r,\theta,\phi)$:
\beq \label{zerotimerho}
\rho(r,\theta,\phi)=\rho_0(r)+\delta\rho_B(r,\theta)+\delta\rho_\alpha(r,\theta,\phi) .
\eeq
Note that in writing down the density field in this way, we are using spherical polar coordinates orientated such that the $z$-axis coincides with the assumed axis of axisymmetry of the magnetic field.  

The idea is that such a star will precess, if the magnetic and rotational and magnetic axes are misaligned, as described in some detail in  \citet{mestel1}.  One can use axes that rotate with the star, such that the $z$-axis remains always along the magnetic axis, while the rotation axis is observed to move in a cone about the $z$-axis, at the (slow) precession frequency.  

The effective ellipticity of such a star comes entirely from the magnetic deformation, such that this slow precession has frequency
\beq \label{omega}
\omega = \alpha\cos\chi\ \frac{I_{zz}-I_{xx}}{I_0}
     \equiv \alpha\epsilon_B\cos\chi,
\eeq
where $\epsilon_B$ is the magnetically-induced ellipticity, and $\chi$ is the angle between the (invariant) angular momentum axis and the magnetic axis.  This angle is sometimes refereed to as the \emph{inclination angle}, or, in this particular context, the \emph{wobble angle}.  Note that we only consider very young stars, before the onset of crust solidification, so we do not allow for any elastic component in the moment of inertia tensor.

\begin{figure}[h]
  \centerline{\includegraphics[width=10cm]{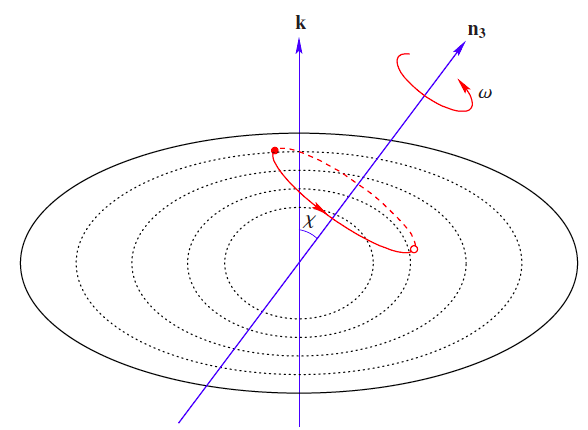}}
  \caption{Schematic illustration of how a fluid element moves in a circle, as described by the base of a cone of half-angle $\chi$, about the magnetic axis $\bf k$.   Reproduced from \citet{LJ17}, `Non-rigid precession of magnetic stars', left hand side of Figure (2).  \label{fig:alphaframe}}
\end{figure}

It follows that the centrifugal bulge itself moves in a cone around the magnetic axis at the slow precession frequency.  This is represented schematically in Figure \ref{fig:alphaframe}.  This means that the total velocity field of a fluid element is the sum of the motion one would expect for a classical rigid body, plus an extra piece due to this effective density wave propagating around the star.  We can therefore write the velocity of a fluid element as:
\beq
{\bf V}_{\mathrm{inertial}} 
  = \alpha\be_z^{(\alpha)}\times\br + \omega\be_z^{(B)}\times\br  + \bxi 
\eeq
The classical rigid body motion is the sum of the rotation, at a rate $\alpha$, about the fixed angular momentum axis, plus the slow precessional motion at rate $\omega$ about the (instantaneous) location of this axis.  The extra `non-rigid' motions are encoded in the so-called $\xi$-motions, named after the symbol originally used to denote them in \citet{mestel1}.  It is these motions that we wish to compute, as they must play a key role in damping the precessional motion.

\psfrag{om}{$\omega$}
\psfrag{chi}{$\chi$}
\psfrag{n3}{$\bf n_3$}
\psfrag{k}{$\bf k$}

Mestel and collaborators noted that one can without too much difficult compute the density perturbation associated with the motion of the centrifugal bulge.  Then one can write down the equation of mass conservation (equation  (\ref{eq:continuity}) below) to try to compute the velocity field associated with these motions.  This however only provides one equation for three unknowns (the three components of the displacement/velocity field).  This was noted as a problem, and a suggestion made that beyond-linear-order perturbation theory might be relevant, but little progress was subsequently made in solving it.

The resolution to this problem is provided in \citep{LJ17}.  The key was to first assemble the full set of equations describing this magnetic precession problem.  These consist of the Euler equation:
\beq \label{general_Euler}
\frac{\partial{\bf v}}{\partial t} + ({\bf v} \cdot \nabla) {\bf v} + 2 {\mathbf \Omega} \times {\bf v} 
 + \frac{\rmd{\mathbf{\Omega}}}{\rmd t} \times {\bf r}  +  {\mathbf \Omega} \times ({\mathbf \Omega} \times {\bf r})
=- \nabla H - \nabla \Phi + \frac{1}{4\pi\rho} (\nabla \times {\bf B}) \times {\bf B},
\eeq
the definition of the enthalpy:
\begin{equation}
H(\br) \equiv \int_{\hat{P}=0}^{\hat{P} = P(\br)} \frac{\rmd \hat{P}}{\rho(\hat{P})} ,
\end{equation}
the equation of mass conservation:
\beq
\label{eq:continuity}
\frac{\partial \rho}{\partial t} = -\nabla \cdot (\rho{\bf v}) ,
\eeq
Poisson's equation for the gravitational field:
\beq
\nabla^2\Phi = 4\pi G \rho , 
\eeq
the time evolution equation for a magnetic field in a perfectly conducting fluid:
\beq
\frac{\partial {\bf B}}{\partial t} = \nabla \times ({\bf v} \times {\bf B}) ,
\eeq
an equation of state:
\beq
H = H(\rho),
\eeq
and of course the solenoidal constraint on the magnetic field:
\beq
\div\bB = 0 .
\eeq
In these equations, $\bf v$ is the fluid velocity, $\mathbf \Omega$ the rotation rate, $\bf r$ the position vector of the fluid element under consideration, $H$ the enthalpy, $\Phi$ the gravitational potential, $\bf B$ the magnetic field strength, and $\rho$ the density.


We would like a solution to these equations that represents our precessing magnetised star.  This is clearly a formidable problem to solve.  The key lies in recognising that there are \emph{two} small parameters that we can exploit when using perturbation theory.  One measures the rotational deformation of the star:
\beq
\epsilon_\alpha \sim \frac{\Omega^2 R_*^3}{GM} \sim 0.21 \frac{R_6^3 f_{\rm kHz}^2}{M_{1.4}} ,  
\eeq
where $f_{\rm kHz} = f/$kHz.  The other small parameter is related to the size of the magnetic deformation:
\beq
\epsilon_B \sim \frac{B^2 R_*^4}{GM^2} \sim 1.9 \times 10^{-6} \frac{B_{15}^2 R_6^4}{M_{1.4}^2}  ,
\label{epsB_scaling}
\eeq
where $B_{15} = B / 10^{15}$ G, $M_{1.4} = M / 1.4 M_\odot$, and $R_6 = R / 10^6$ cm.  Clearly, these will both be small for realistic rotation rates and magnetic field strengths. 

This immediately suggests a $2$-parameter perturbative scheme.  In an obvious notation we can expand the various quantities that appear in our equations.  For instance, for the density we can write:
\begin{equation} 
\rho = \rho_0 + \delta \rho_{\alpha} + \delta \rho_{B} + \delta \rho_{\alpha \alpha} + \delta \rho_{\alpha B} + \delta \rho_{BB} + \dots 
\end{equation}
For the magnetic field we can write:
\begin{equation}
{\bf B} = \bB_0 + \delta \bB_\alpha + \delta\bB_B + \dots 
\end{equation}
We can then assemble coupled sets of equations, order by order.

To zeroth order we have:
\begin{equation}
\label{eq:zeroth_Euler}
0 = -\nabla H_0 - \nabla \Phi_0 ,
\end{equation}
\begin{equation}
\nabla^2 \Phi_0 = 4\pi G \rho_0 ,
\end{equation}
\begin{equation}
H_0 = H_0(\rho_0) .
\end{equation}
This simply represents a static spherical star.  If we specialise to an $n=1$ polytrope for the equation of state, we have the well known density profile:
\beq \label{rho_0}
\rho_0=\rho_0(r)=\rho_c\frac{R_*\sin\brac{\Mr}}{\pi r} .
\eeq

To first order in $\epsilon_B$ we have:
\begin{equation}
0 = -\nabla \delta H_B - \nabla \delta  \Phi_B + \frac{1}{4 \pi \rho_0} (\nabla \times \bB_0) \times \bB_0  ,
\end{equation}
\begin{equation}
\nabla^2 \delta \Phi_B = 4\pi G \delta \rho_B ,
\end{equation}
\begin{equation}
\delta H_B = \td{H}{\rho} \delta\rho_B ,
\end{equation}
\begin{equation}
\nabla \cdot \bB_0 = 0 .
\end{equation}
This represents a static star deformed by a static magnetic field.  Using techniques described in \citet{LJ09} one can find a solution:
\beq \label{B0}
\bB_0=B_\phi\be_\phi = \Lambda\rho_0 r\sin\theta\be_\phi,
\eeq
where the $\Lambda$ is a coefficient that can be determined for a given star and a given magnetic field configuration.  We consider toroidal magnetic fields here, so as to give prolate deformations.  For the field employed here we have 
\beq \label{epsB}
\epsilon_B = k_B \frac{\Lambda^2}{G} \approx -0.019 \frac{\Lambda^2}{G},
\eeq
where the coefficient that appears comes from a full numerical solution of the equations \citet{LJ09}.

To first order in $\epsilon_\alpha$ we have:
\begin{equation}
\label{eq:alpha_Euler}
 {\mathbf \Omega}_\alpha \times ({\mathbf \Omega}_\alpha \times {\bf r}) 
 =  -\nabla \delta H_\alpha - \nabla \delta  \Phi_\alpha  ,
\end{equation}
\begin{equation}
\nabla^2 \delta \Phi_\alpha = 4\pi G \delta \rho_\alpha ,
\end{equation}
\begin{equation}
\delta H_\alpha = \td{H}{\rho} \delta \rho_\alpha .
\end{equation}
This represents a star deformed by rotation, with the complication that the rotation axis moves slowly relative to the star.  At this level of approximation, the solution is simply a `snap-shot', i.e. given by the density field appropriate to the instantaneous  location of the spin axis relative to the star.  Explicitly:
\beq \label{delta_rho_alpha}
\delta\rho_\alpha
 = -\frac{5\pi\alpha^2}{16 G}
       j_2\brac{\Mr}
       \brac{ \sin(2\chi)\sin(2\theta)\cos\lambda + \sin^2\chi\sin^2\theta\cos(2\lambda) },
\eeq
where $\lambda = \phi + \omega t$.

If we work to order $\epsilon_\alpha \epsilon_B$ we have:
\begin{equation} \nonumber
\frac{\rmd{\mathbf \Omega_\alpha}}{\rmd t} \times {\bf r} 
+
\omega \be_z \times ({\mathbf \Omega_\alpha} \times {\bf r})
+
{\mathbf \Omega_\alpha}  \times (\omega \be_z  \times {\bf r})
 =
 -\nabla \delta H_{\alpha B}  - \nabla \delta  \Phi_{\alpha B}
 - \frac{\delta\rho_\alpha}{4\pi \rho_0^2} (\nabla \times \bB_0) \times \bB_0 
 \end{equation}
\begin{equation}   \label{Euler_aB}
\hspace{60mm}  
+  \frac{1}{4\pi \rho_0} [ (\nabla \times \delta \bB_\alpha) \times \bB_0  + (\nabla \times \bB_0) \times \delta\bB_\alpha ]   ,
\end{equation}

\begin{equation}
\nabla^2 \delta \Phi_{\alpha B} = 4\pi G \delta \rho_{\alpha B} ,
\end{equation}

\begin{equation}
\delta H_{\alpha B} = \td{H}{\rho} \delta\rho_{\alpha B} + \td{^2 H}{\rho^2} \delta \rho_{\alpha}\delta \rho_B  ,
\end{equation}

\begin{equation} 
\label{cont_aB}
\frac{\partial \delta \rho_\alpha}{\partial t} + \nabla \cdot (\rho_0 {\boldmath \dot\xi}) = 0 ,
\end{equation}

\begin{equation} \label{induct_aB}
\frac{\partial\delta\bB_\alpha}{\partial t} - \nabla \times ({\boldmath \dot\xi} \times \bB_0) = 0 .
\end{equation}
This constitutes a set of nine equations in the nine unknowns $(\delta \rho_{\alpha B}, \delta H_{\alpha B}, \delta \Phi_{\alpha B},{\boldmath \dot\xi}, \delta\bB_\alpha$).   The set represents the combined effect of rotation and magnetic fields, and, crucially, allows computation of the all-important $\xi$-motions.

\begin{figure}[h]
  \centerline{\includegraphics[width=15cm]{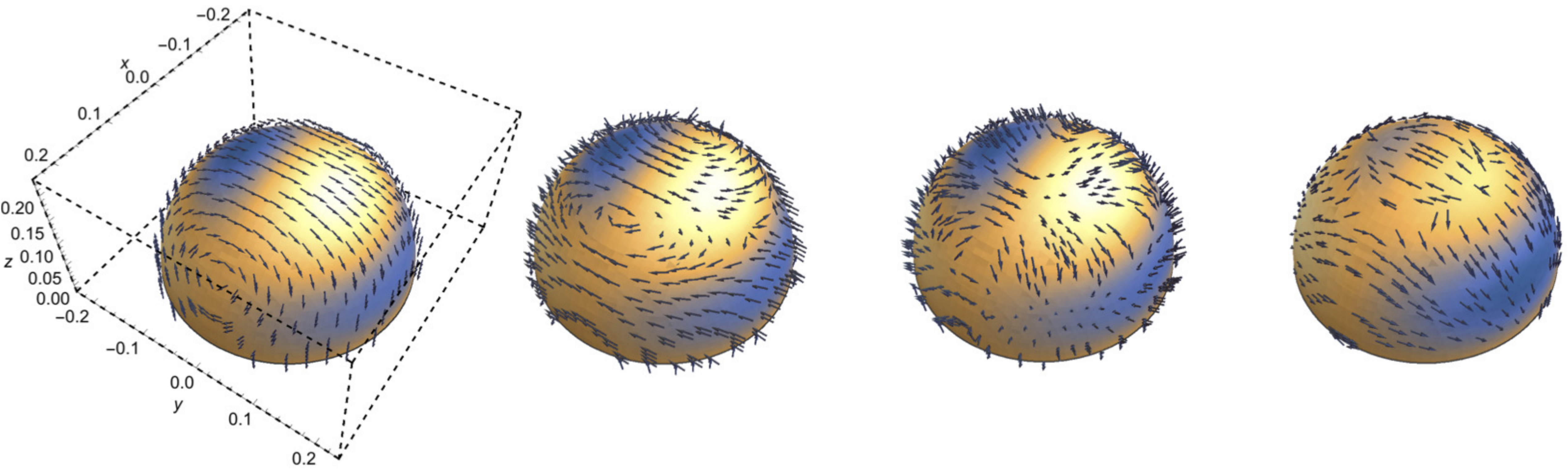}}
  \caption{Illustration of the displacement field $\xi$ on four hemispherical shells, corresponding to radii of $0.2, 0.4, 0.6, 0.8$ of the total stellar radius, as read from left to right. Reproduced from \citet{LJ17}, `Non-rigid precession of magnetic stars', Figure (8).   \label{fig:xi-motion}}
\end{figure}

A method of solution is set out in detail in \citet{LJ17}, where the solution to the problem is written in terms of spherical harmonics.  A solution retaining harmonics up to and including $l= 4$-terms is given in their section 7.1.  A graphical representation is reproduced in Figure \ref{fig:xi-motion}.  It is this displacement field that can now be used to investigate dissipation of the precessional motion.

\section{Formalism for evolving the spin vector}\label{sect:ffetsv}

We now need to write down the equations that will allow us to see how the precessional motion evolves in time, when electromagnetic torques and internal dissipation are added.

The magnetic axis spins around the angular momentum axis at a rate 
\beq
\label{eq:J_phi_dot} 
\dot\phi = \frac{J}{I_1} ,
\eeq
which can be thought of as the magnitude of the spin rate.

We know that the star's energy will decrease, due to both electromagnetic torque, and viscous dissipation:
\beq
\label{eq:E_dot_total}
\dot E = \dot E_{\rm visc} + \dot E_{\rm EM} ,
\eeq 
while the angular momentum decreases only by virtue of the electromagnetic torques:
\beq
\dot J = \dot J_{\rm EM}.
\eeq
The star's energy will be a function of both the angular momentum and the wobble angle, so we can write:
\beq
\dot E = \frac{\partial E}{\partial J} \dot J + \frac{\partial E}{\partial \chi} \dot\chi .
\eeq
This can be rearranged to give us an evolution equation for $\chi$:
\beq
\dot\chi = \frac{\dot E - \frac{\partial E}{\partial J} \dot J}{\frac{\partial E}{\partial \chi}} .
\eeq
If we consider spin-down by a vacuum dipole, one can show (see discussion in \citet{cutler_jones_01} for the analogous gravitational wave torque case):
\beq 
\label{eq:E_dot_OG}
\dot E_{\rm EM} = \dot\phi \dot J_{\rm EM} .
\eeq
We then obtain
\beq \label{chi_dot_general}
\dot\chi = \frac{\dot E_{\rm visc} + 
\dot J_{\rm EM}\left(\dot\phi - \frac{\partial E}{\partial J}\right) }{\frac{\partial E}{\partial \chi}} ,
\eeq
and
\beq
\ddot\phi = \frac{\dot J_{\rm EM}}{I_1},
\eeq
which to our order of working may be rewritten
\beq \label{eq:alpha-dot}
\dot\Omega = \frac{\dot J_{\rm EM}}{I}.
\eeq
We now have two (coupled) evolutions equations for the wobble angle $\chi$ and the spin rate $\Omega$, equations (\ref{chi_dot_general}) and (\ref{eq:alpha-dot}), which can then be solved if we specify the nature of the torques and the damping mechanisms.

\section{Bulk viscosity}\label{sect:bv}

To account for viscous dissipation one needs to add terms involving shear viscosity (with coefficient $\eta$) and bulk viscosity (with coefficient $\zeta$) to the Euler equation, to give:
\beq
\rho\left[  \frac{\partial{\boldsymbol v}}{\partial t} + ({\boldsymbol v} \cdot \nabla) {\boldsymbol v} \right] 
=- \nabla P - \rho\nabla \Phi + \frac{1}{4\pi} (\nabla \times {\boldsymbol B}) \times {\boldsymbol B}\nn
+2\div(\eta{\boldsymbol{\sigma}})+\nabla(\zeta\div\bv),
\eeq
where $\sigma_{ab}$ measures the amount of shear in the motion:
\beq
\sigma_{ab}=\frac{1}{2}(\nabla_a v_b+\nabla_b v_a)-\frac{1}{3}\nabla_c v_c g_{ab} ;
\eeq
see e.g. \citet{lindblom_owen_02}.

Back-of-the-envelope estimates readily show that in the regime of interest to us, bulk viscosity is much more important than shear, so we will neglect shear dissipation; see \citet{LJ18}.  Also, the damping timescales are long compared with the free precession period, so we can use the dissipation-free eigenfunctions computed above to estimate the rate of energy dissipation.

The bulk viscosity coefficient itself takes the form
\begin{equation}
\label{eq:zeta_general}
\zeta \approx - \frac{n \tau}{1 + (\omega \tau)^2} \left.\frac{\partial P}{\partial x}\right|_n \frac{\rmd x}{\rmd n} .
\end{equation}
In this equation, $\omega$ is the (angular) frequency of the free precession, while $\tau$ is the timescale of the microphysical relaxation process that supplies the bulk viscosity.  Note that at fixed $\omega$, this coefficient has a maximum when $\omega = 1 / \tau$, i.e. there is a `resonance'-like feature in the bulk viscosity magnitude.  

We will consider dissipation due to modified Urca reactions which introduces a steep temperature dependence in $\tau$, as described in \citet{reisenegger_goldreich_92}:
\begin{equation}
\label{eq:RG_relaxation_timescale}
\tau \sim \frac{0.2}{T_9^6} \left(\frac{\rho}{\rho_{\rm nuc}}\right)^{2/3} \, {\rm yr} .
\end{equation}
This means that the stellar temperature will be important when attempting to evolve the precessional motion.  We will use a simple formula for cooling, also due to modified Urca reactions, as described in \citet{pgw_06}:
\beq\label{murca_approx}
T(t)=\brac{\frac{6 N^s}{C}t+\frac{1}{T_0^6}}^{-1/6} .
\eeq

\section{The electromagnetic torque}\label{sect:tet}

We will consider the overly-simple case of vacuum dipole spin-down, for which:
\beq
\label{eq:E_dot_EM}
\dot E_{\rm EM}  = \Omega \dot J_{\rm EM} = -\frac{R^6}{6c^3} \Omega^4 B_{\rm p}^2 \lambda(\chi) ,
\eeq
with $\lambda(\chi)$ a function that encodes how the torque varies with $\chi$, with $\lambda(\chi) = \sin^2\chi$ for the vacuum dipole considered here.  Realistic torques will include a magnetospheric component, which will give non-zero spin-down even for aligned $(\chi = 0)$ stars.  We will discuss this point in the Section `Future Work' below.

\section{Time evolutions}\label{sect:tw}

We have now assembled all the ingredients needed to evolve the quantities $\Omega, \chi$, with the temperature evolution being given by equation (\ref{murca_approx}).  Before presenting the results of such evolutions, we can gain a little insight as follows.

\begin{figure}
\begin{minipage}[c]{\linewidth}
\includegraphics[width=\textwidth]{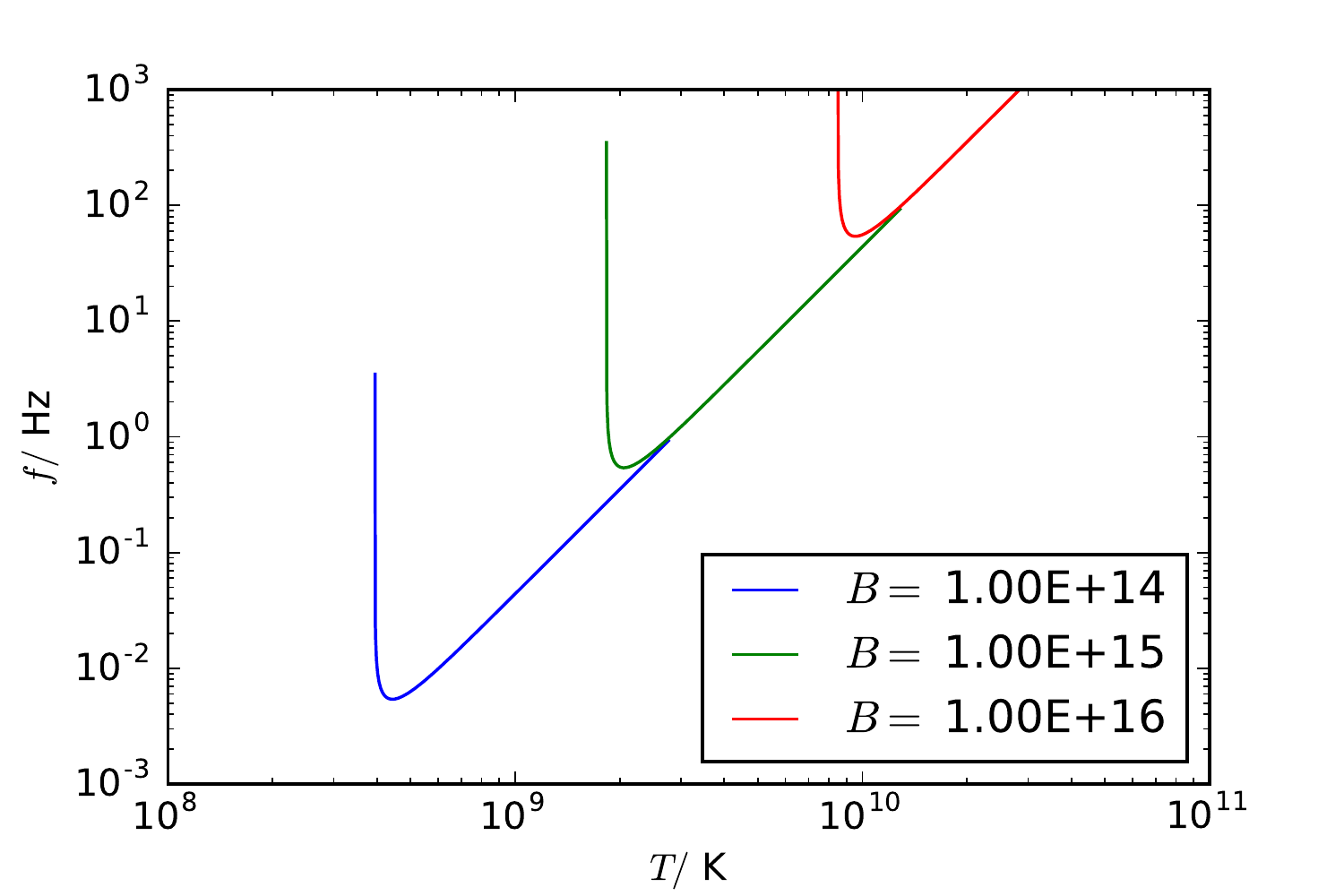}   
\caption{  Critical curve for orthogonalisation.  Stars above the curve tend to orthogonalise, while stars below the curve tend to align.  Three such curves are plotted, for the magnetic field strength indicated. Reproduced from \citet{LJ18}, `Neutron-star spindown and magnetic inclination-angle evolution', Figure (4).   \label{fig:f_crit_v_T_multi}}
\end{minipage}
\end{figure}

The electromagnetic torque tends to make the star align.  For a prolate star, the internal dissipation tends to make it orthogonalise.  It follows that these two effects will balance when their respective timescales of operation are equal: 
\begin{equation}
\tau_{\chi}^{\rm EM} = \tau_{\chi}^{\rm bulk} ,
\end{equation}
where we have defined a timescale of the form
\begin{equation}
\tau_\chi \equiv \frac{\sin\chi}{\frac{d}{dt} \sin\chi} .
\end{equation}
By combing equations given above, it can be shown that this occurs when the spin frequency takes a critical value:
\begin{equation} \label{fcrit}
f^{\rm crit}_{\rm kHz} \approx \frac{2.8 T_{10}^6}{[4.1 \times 10^3 \tilde\lambda(\chi) T_{10}^6 - 0.155 B_{15}^4 \cos^2\chi]^{1/2} },
\end{equation}
where for convenience we have defined
\begin{equation}
\tilde\lambda(\chi) = \frac{g(\chi)}{\cos^2\chi \lambda(\chi)} ,
\end{equation}
where $g(\chi)$ is a dimensionless function that encodes how the rate of viscous dissipation varies with $\chi$.  For small $\chi$, it can be shown that $g(\chi) \sim \sin^2\chi \sim \chi^2$; see \citet{LJ18} for details.  Stars with spin frequencies above this tend to orthogonalise, those with frequencies below this tend to align.  This equation therefore defines a critical curve for orthogonalisation in the spin frequency--temperature plane.  A family of such curves, for three different magnetic field strengths, is given in Figure \ref{fig:f_crit_v_T_multi}, where the $\chi$-dependent factors have been set equal to unity.  
One can use such curves to anticipate how a star may evolve.  

\begin{figure}
\begin{minipage}[c]{\linewidth}
\includegraphics[width=\textwidth]{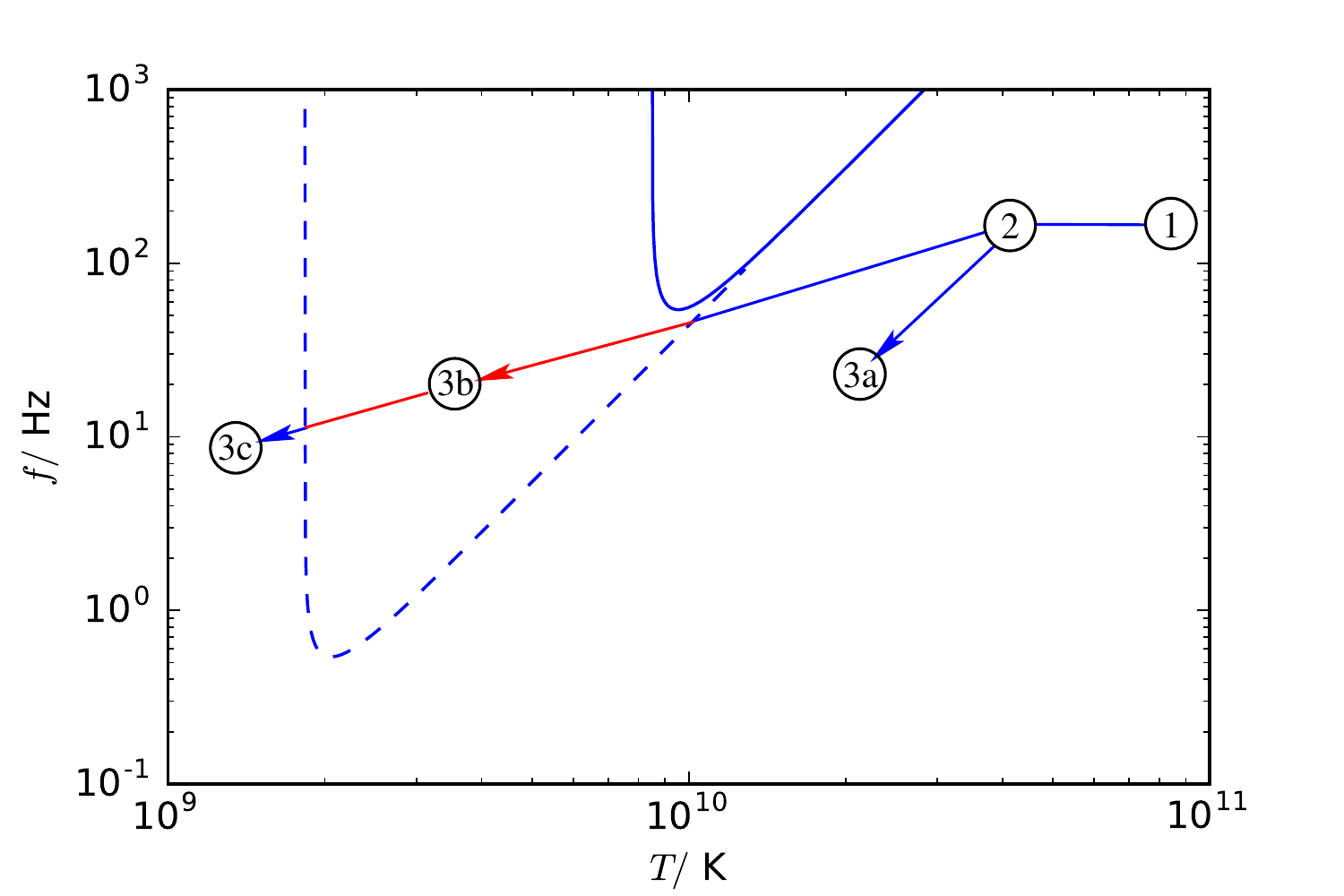}   
\caption{  Some schematic trajectories that a star might follow in the spin frequency--temperature plane.   Reproduced from \citet{LJ18}, `Neutron-star spindown and magnetic inclination-angle evolution', Figure (5).   \label{track_in_window}}
\end{minipage}
\end{figure}

Consider the possible trajectories sketched, again schematically, in Figure \ref{track_in_window}.  Stars will be born hot and spinning fast, somewhere in the top right of the plot, marked by position 1.  The star will them cool rapidly, moving left in the plot, to position 2.  If the star spins down significantly it may fail to enter the orthogonalisation region, and so never othogonalise.  This is the case for the trajectory that ends at the point 3a, relevant for a high field strength orthogonalisation curve.  However, for most weakly magnetised stars, the larger dotted orthogonalisation curve applies.  In this case the star may enter and go through a period of orthogonalisation, denoted by 3b in the plot.  It will eventually cool through the orthogonalisation region, exiting to evolve to 3c, where the lower temperatures and slower rotation rate will make subsequent evolution slow.  Such a star may be driven to orthogonality.

\begin{figure}
\begin{minipage}[c]{\linewidth}
\includegraphics[width=15cm]{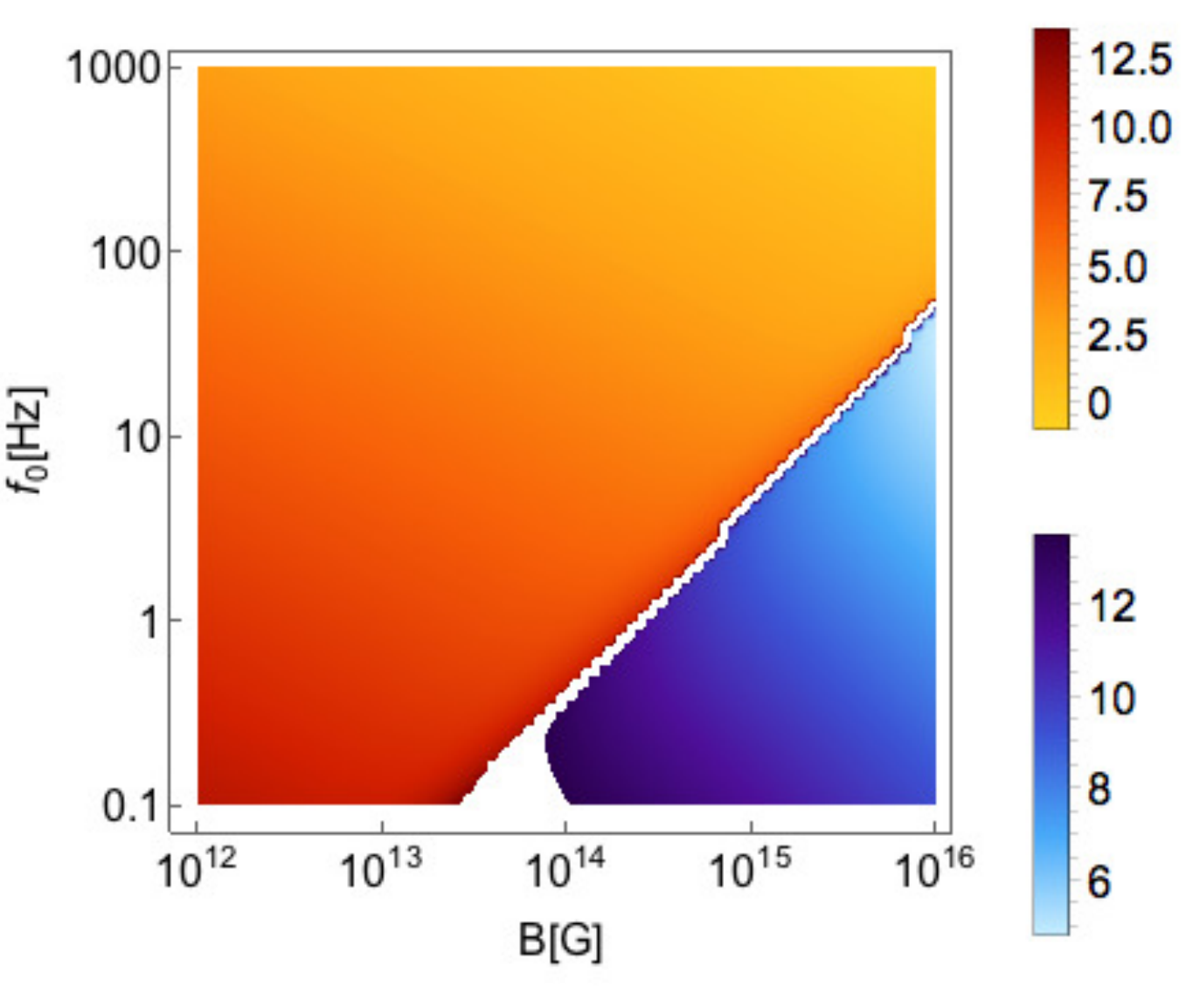}     
\caption{  Colour scale plot show the time taken for stars to either orthogonalise (orange) or align (blue), as a function of location in the birth spin frequency--magnetic field strength plane.   Reproduced from \citet{LJ18}, `Neutron-star spindown and magnetic inclination-angle evolution', left hand side of Figure (14).    \label{times_vac_vs_spit}}
\end{minipage}
\end{figure}

The results of actual time evolutions are shown in Figure \ref{times_vac_vs_spit}.  This shows the phase space of birth spin frequency and magnetic field strength.  The colour scale is used to indicate the nature of the stellar configuration after various times since birth.  The blue colour scale shows the time taken (in log seconds) for states to align, for those stars that do indeed align.  Orange is used to show the time taken for stars to orthogonalise.  Note that we evolve to timescales of order a million years.  However, we only trust our model prior to the formation of the crust, which occurs on a timescale of a hundred years or less.  We show such long term evolutions simply to make clearer the nature of the solutions.

We can summarise the main features as follows.  Below a field strength of about $10^{14}$ G, all stars orthogonalise, regardless of birth spin frequency.  This corresponds to the orthogonalisation region of Figure \ref{track_in_window} being large, so that all stars enter and undergo a significant period of orthogonalisation.  Above $10^{14}$ G, there is a (field-dependent) spin frequency below which stars align, and above which star orthogonalise.  This can be understood again in terms of Figure \ref{track_in_window}: only those stars that are born spinning fast enough can enter the orthogonalisation region, and undergo a significant period of orthogonalisation.

\section{Future work}\label{sect:fw}

The above results illustrate the interplay of internal dissipation and electromagnetic torques that play key roles in the early life of a millisecond magnetar.  However, the model is highly idealised, and a number of issues will need to be addressed before firm conclusions can be drawn.  We will now discuss the most important outstanding issues.

\emph{The form of the spin-down torque}.  The model and results above assume a vacuum dipole model for the electromagnetic spin-down torque.  However, it has been known for a long time that a significant fraction of the star's angular momentum is likely to be carried away by particles accelerated within the pulsar's magnetosphere \citep{GJ_69}.  In fact, in the paper \cite{LJ18} we presented results for such a torque, with angular dependence $\lambda(\chi) = 1 +\sin^2\chi$.  Such a torque is qualitatively different from the vacuum dipole one, as in the limit $\chi \rightarrow 0$ the vacuum torque goes to zero, while the magnetospheric one does not.  However, there was an inconsistency in our previous treatment, as we employed equation (\ref{eq:E_dot_OG}),which is valid for the vacuum dipole torque, but not for the magnetospheric one.  We aim to correct this in a future publication.

\emph{Neglect of buoyancy}. In perturbing the Euler equation above, we have implicitly assumed that the perturbations obey the same equation of state as the background (unperturbed) star.  In general this will not be the case.  In particular, when the reactions responsible for the bulk viscosity are slow compared with the mode period, we might expect the eigenfunctions themselves to be different from those calculated in \citet{LJ17}, with a smaller divergence $\nabla \cdot \xi$, as described in \citet{lasky_glam}.  This could be modelled by shutting off the effects of bulk viscosity when the reaction timescales becomes sufficiently long, as was done in \citet{lasky_glam}.  

\emph{Opacity to neutrinos}.  When the star is sufficiently hot, it will be opaque to neutrinos.   This could be accounted for in future work using the prescription of \citet{lasky_glam}, who switched bulk viscosity off above a critical temperature.

\emph{Evolution of the temperature}.  Our prescription for evolving the temperature was particularly simple.  In reality, additional cooling effects may be important in the early life of magnetars, as described in \citet{metzger}.  

We intend addressing all of these issues in future work (Lander \& Jones, in preparation).  In the meantime, we feel the scheme here, with its competition between electromagnetic alignment and dissipative orthogonalisation, has taken an important step forward in modelling the early life of a millisecond magnetar.

\section{Acknowledgement}

DIJ acknowledges funding from STFC through grant number ST/M000931/1.


\bibliographystyle{aipnum-cp}

\bibliography{references}

\begin{thebibliography}{14}%
\makeatletter
\providecommand \@ifxundefined [1]{%
 \@ifx{#1\undefined}
}%
\providecommand \@ifnum [1]{%
 \ifnum #1\expandafter \@firstoftwo
 \else \expandafter \@secondoftwo
 \fi
}%
\providecommand \@ifx [1]{%
 \ifx #1\expandafter \@firstoftwo
 \else \expandafter \@secondoftwo
 \fi
}%
\providecommand \natexlab [1]{#1}%
\providecommand \enquote  [1]{``#1''}%
\providecommand \bibnamefont  [1]{#1}%
\providecommand \bibfnamefont [1]{#1}%
\providecommand \citenamefont [1]{#1}%
\providecommand \href@noop [0]{\@secondoftwo}%
\providecommand \href [0]{\begingroup \@sanitize@url \@href}%
\providecommand \@href[1]{\@@startlink{#1}\@@href}%
\providecommand \@@href[1]{\endgroup#1\@@endlink}%
\providecommand \@sanitize@url [0]{\catcode `\$12\catcode `\&12\catcode
  `\#12\catcode `\^12\catcode `\_12\catcode `\%12\relax}%
\providecommand \@@startlink[1]{}%
\providecommand \@@endlink[0]{}%
\providecommand \url  [0]{\begingroup\@sanitize@url \@url }%
\providecommand \@url [1]{\endgroup\@href {#1}{\urlprefix }}%
\providecommand \urlprefix  [0]{URL }%
\providecommand \Eprint [0]{\href }%
\providecommand \doibase [0]{http://dx.doi.org/}%
\providecommand \selectlanguage [0]{\@gobble}%
\providecommand \bibinfo  [0]{\@secondoftwo}%
\providecommand \bibfield  [0]{\@secondoftwo}%
\providecommand \translation [1]{[#1]}%
\providecommand \BibitemOpen [0]{}%
\providecommand \bibitemStop [0]{}%
\providecommand \bibitemNoStop [0]{.\EOS\space}%
\providecommand \EOS [0]{\spacefactor3000\relax}%
\providecommand \BibitemShut  [1]{\csname bibitem#1\endcsname}%
\let\auto@bib@innerbib\@empty
\bibitem [{\citenamefont {{Lander}}\ and\ \citenamefont
  {{Jones}}(2017)}]{LJ17}%
  \BibitemOpen
  \bibfield  {author} {\bibinfo {author} {\bibfnamefont {S.~K.}\ \bibnamefont
  {{Lander}}}\ and\ \bibinfo {author} {\bibfnamefont {D.~I.}\ \bibnamefont
  {{Jones}}},\ }\href {\doibase 10.1093/mnras/stx349} {\bibfield  {journal}
  {\bibinfo  {journal} {\mnras}\ }\textbf {\bibinfo {volume} {467}},\ \unskip\
  \bibinfo {pages} {4343--4382}June (\bibinfo {year} {2017})},\ \Eprint
  {http://arxiv.org/abs/1610.08745} {arXiv:1610.08745 [astro-ph.SR]}
  \BibitemShut {NoStop}%
\bibitem [{\citenamefont {{Lander}}\ and\ \citenamefont
  {{Jones}}(2018)}]{LJ18}%
  \BibitemOpen
  \bibfield  {author} {\bibinfo {author} {\bibfnamefont {S.~K.}\ \bibnamefont
  {{Lander}}}\ and\ \bibinfo {author} {\bibfnamefont {D.~I.}\ \bibnamefont
  {{Jones}}},\ }\href {\doibase 10.1093/mnras/sty2553} {\bibfield  {journal}
  {\bibinfo  {journal} {\mnras}\ }\textbf {\bibinfo {volume} {481}},\ \unskip\
  \bibinfo {pages} {4169--4193}December (\bibinfo {year} {2018})},\ \Eprint
  {http://arxiv.org/abs/1807.01289} {arXiv:1807.01289 [astro-ph.HE]}
  \BibitemShut {NoStop}%
\bibitem [{\citenamefont {{Dall'Osso}}, \citenamefont {{Shore}},\ and\
  \citenamefont {{Stella}}(2009)}]{dallosso09}%
  \BibitemOpen
  \bibfield  {author} {\bibinfo {author} {\bibfnamefont {S.}~\bibnamefont
  {{Dall'Osso}}}, \bibinfo {author} {\bibfnamefont {S.~N.}\ \bibnamefont
  {{Shore}}}, \ and\ \bibinfo {author} {\bibfnamefont {L.}~\bibnamefont
  {{Stella}}},\ }\href {\doibase 10.1111/j.1365-2966.2008.14054.x} {\bibfield
  {journal} {\bibinfo  {journal} {\mnras}\ }\textbf {\bibinfo {volume} {398}},\
  \unskip\ \bibinfo {pages} {1869--1885}October (\bibinfo {year} {2009})},\
  \Eprint {http://arxiv.org/abs/0811.4311} {arXiv:0811.4311} \BibitemShut
  {NoStop}%
\bibitem [{\citenamefont {{Dall'Osso}}, \citenamefont {{Stella}},\ and\
  \citenamefont {{Palomba}}(2018)}]{dallosso_spin}%
  \BibitemOpen
  \bibfield  {author} {\bibinfo {author} {\bibfnamefont {S.}~\bibnamefont
  {{Dall'Osso}}}, \bibinfo {author} {\bibfnamefont {L.}~\bibnamefont
  {{Stella}}}, \ and\ \bibinfo {author} {\bibfnamefont {C.}~\bibnamefont
  {{Palomba}}},\ }\href@noop {} {\bibfield  {journal} {\bibinfo  {journal}
  {ArXiv e-prints}\ June} (\bibinfo {year} {2018})},\ \Eprint
  {http://arxiv.org/abs/1806.11164} {arXiv:1806.11164 [astro-ph.HE]}
  \BibitemShut {NoStop}%
\bibitem [{\citenamefont {{Cutler}}(2002)}]{cutler_02}%
  \BibitemOpen
  \bibfield  {author} {\bibinfo {author} {\bibfnamefont {C.}~\bibnamefont
  {{Cutler}}},\ }\href {\doibase 10.1103/PhysRevD.66.084025} {\bibfield
  {journal} {\bibinfo  {journal} {\prd}\ }\textbf {\bibinfo {volume} {66}},\
  p.\ \bibinfo {pages} {084025}October (\bibinfo {year} {2002})},\ \Eprint
  {http://arxiv.org/abs/gr-qc/0206051} {gr-qc/0206051} \BibitemShut {NoStop}%
\bibitem [{\citenamefont {{Metzger}}\ \emph {et~al.}(2011)\citenamefont
  {{Metzger}}, \citenamefont {{Giannios}}, \citenamefont {{Thompson}},
  \citenamefont {{Bucciantini}},\ and\ \citenamefont {{Quataert}}}]{metzger}%
  \BibitemOpen
  \bibfield  {author} {\bibinfo {author} {\bibfnamefont {B.~D.}\ \bibnamefont
  {{Metzger}}}, \bibinfo {author} {\bibfnamefont {D.}~\bibnamefont
  {{Giannios}}}, \bibinfo {author} {\bibfnamefont {T.~A.}\ \bibnamefont
  {{Thompson}}}, \bibinfo {author} {\bibfnamefont {N.}~\bibnamefont
  {{Bucciantini}}}, \ and\ \bibinfo {author} {\bibfnamefont {E.}~\bibnamefont
  {{Quataert}}},\ }\href {\doibase 10.1111/j.1365-2966.2011.18280.x} {\bibfield
   {journal} {\bibinfo  {journal} {\mnras}\ }\textbf {\bibinfo {volume}
  {413}},\ \unskip\ \bibinfo {pages} {2031--2056}May (\bibinfo {year}
  {2011})},\ \Eprint {http://arxiv.org/abs/1012.0001} {arXiv:1012.0001
  [astro-ph.HE]} \BibitemShut {NoStop}%
\bibitem [{\citenamefont {{Mestel}}\ and\ \citenamefont
  {{Takhar}}(1972)}]{mestel1}%
  \BibitemOpen
  \bibfield  {author} {\bibinfo {author} {\bibfnamefont {L.}~\bibnamefont
  {{Mestel}}}\ and\ \bibinfo {author} {\bibfnamefont {H.~S.}\ \bibnamefont
  {{Takhar}}},\ }\href {\doibase 10.1093/mnras/156.4.419} {\bibfield  {journal}
  {\bibinfo  {journal} {\mnras}\ }\textbf {\bibinfo {volume} {156}},\ p.\
  \bibinfo {pages} {419} (\bibinfo {year} {1972})}\BibitemShut {NoStop}%
\bibitem [{\citenamefont {{Lander}}\ and\ \citenamefont
  {{Jones}}(2009)}]{LJ09}%
  \BibitemOpen
  \bibfield  {author} {\bibinfo {author} {\bibfnamefont {S.~K.}\ \bibnamefont
  {{Lander}}}\ and\ \bibinfo {author} {\bibfnamefont {D.~I.}\ \bibnamefont
  {{Jones}}},\ }\href {\doibase 10.1111/j.1365-2966.2009.14667.x} {\bibfield
  {journal} {\bibinfo  {journal} {\mnras}\ }\textbf {\bibinfo {volume} {395}},\
  \unskip\ \bibinfo {pages} {2162--2176}June (\bibinfo {year} {2009})},\
  \Eprint {http://arxiv.org/abs/0903.0827} {arXiv:0903.0827 [astro-ph.SR]}
  \BibitemShut {NoStop}%
\bibitem [{\citenamefont {{Cutler}}\ and\ \citenamefont
  {{Jones}}(2001)}]{cutler_jones_01}%
  \BibitemOpen
  \bibfield  {author} {\bibinfo {author} {\bibfnamefont {C.}~\bibnamefont
  {{Cutler}}}\ and\ \bibinfo {author} {\bibfnamefont {D.~I.}\ \bibnamefont
  {{Jones}}},\ }\href {\doibase 10.1103/PhysRevD.63.024002} {\bibfield
  {journal} {\bibinfo  {journal} {\prd}\ }\textbf {\bibinfo {volume} {63}},\
  p.\ \bibinfo {pages} {024002}January (\bibinfo {year} {2001})},\ \Eprint
  {http://arxiv.org/abs/gr-qc/0008021} {gr-qc/0008021} \BibitemShut {NoStop}%
\bibitem [{\citenamefont {{Lindblom}}\ and\ \citenamefont
  {{Owen}}(2002)}]{lindblom_owen_02}%
  \BibitemOpen
  \bibfield  {author} {\bibinfo {author} {\bibfnamefont {L.}~\bibnamefont
  {{Lindblom}}}\ and\ \bibinfo {author} {\bibfnamefont {B.~J.}\ \bibnamefont
  {{Owen}}},\ }\href {\doibase 10.1103/PhysRevD.65.063006} {\bibfield
  {journal} {\bibinfo  {journal} {\prd}\ }\textbf {\bibinfo {volume} {65}},\
  p.\ \bibinfo {pages} {063006}March (\bibinfo {year} {2002})},\ \Eprint
  {http://arxiv.org/abs/astro-ph/0110558} {astro-ph/0110558} \BibitemShut
  {NoStop}%
\bibitem [{\citenamefont {{Reisenegger}}\ and\ \citenamefont
  {{Goldreich}}(1992)}]{reisenegger_goldreich_92}%
  \BibitemOpen
  \bibfield  {author} {\bibinfo {author} {\bibfnamefont {A.}~\bibnamefont
  {{Reisenegger}}}\ and\ \bibinfo {author} {\bibfnamefont {P.}~\bibnamefont
  {{Goldreich}}},\ }\href {\doibase 10.1086/171645} {\bibfield  {journal}
  {\bibinfo  {journal} {\apj}\ }\textbf {\bibinfo {volume} {395}},\ \unskip\
  \bibinfo {pages} {240--249}August (\bibinfo {year} {1992})}\BibitemShut
  {NoStop}%
\bibitem [{\citenamefont {{Page}}, \citenamefont {{Geppert}},\ and\
  \citenamefont {{Weber}}(2006)}]{pgw_06}%
  \BibitemOpen
  \bibfield  {author} {\bibinfo {author} {\bibfnamefont {D.}~\bibnamefont
  {{Page}}}, \bibinfo {author} {\bibfnamefont {U.}~\bibnamefont {{Geppert}}}, \
  and\ \bibinfo {author} {\bibfnamefont {F.}~\bibnamefont {{Weber}}},\ }\href
  {\doibase 10.1016/j.nuclphysa.2005.09.019} {\bibfield  {journal} {\bibinfo
  {journal} {Nuclear Physics A}\ }\textbf {\bibinfo {volume} {777}},\ \unskip\
  \bibinfo {pages} {497--530}October (\bibinfo {year} {2006})},\ \Eprint
  {http://arxiv.org/abs/astro-ph/0508056} {astro-ph/0508056} \BibitemShut
  {NoStop}%
\bibitem [{\citenamefont {{Goldreich}}\ and\ \citenamefont
  {{Julian}}(1969)}]{GJ_69}%
  \BibitemOpen
  \bibfield  {author} {\bibinfo {author} {\bibfnamefont {P.}~\bibnamefont
  {{Goldreich}}}\ and\ \bibinfo {author} {\bibfnamefont {W.~H.}\ \bibnamefont
  {{Julian}}},\ }\href {\doibase 10.1086/150119} {\bibfield  {journal}
  {\bibinfo  {journal} {\apj}\ }\textbf {\bibinfo {volume} {157}},\ p.\
  \bibinfo {pages} {869}August (\bibinfo {year} {1969})}\BibitemShut {NoStop}%
\bibitem [{\citenamefont {{Lasky}}\ and\ \citenamefont
  {{Glampedakis}}(2016)}]{lasky_glam}%
  \BibitemOpen
  \bibfield  {author} {\bibinfo {author} {\bibfnamefont {P.~D.}\ \bibnamefont
  {{Lasky}}}\ and\ \bibinfo {author} {\bibfnamefont {K.}~\bibnamefont
  {{Glampedakis}}},\ }\href {\doibase 10.1093/mnras/stw435} {\bibfield
  {journal} {\bibinfo  {journal} {\mnras}\ }\textbf {\bibinfo {volume} {458}},\
  \unskip\ \bibinfo {pages} {1660--1670}May (\bibinfo {year} {2016})},\ \Eprint
  {http://arxiv.org/abs/1512.05368} {arXiv:1512.05368 [astro-ph.HE]}
  \BibitemShut {NoStop}%
\end{thebibliography}%

\end{document}